\documentclass{aastex631}

\usepackage{wrapfig}
\usepackage{gensymb}

\accepted{September 17, 2022}

\shorttitle{nacl \& kcl in io's atmosphere}
\graphicspath{{./}{figures/}}

\begin{document}

\title{NaCl \& KCl in Io's Atmosphere}

\author{Erin Redwing}
\affiliation{University of California - Berkeley, Department of Earth \& Planetary Science, 307 McCone Hall, Berkeley, CA, USA}

\author{Imke de Pater}
\affiliation{University of California - Berkeley, Department of Astronomy, 501 Campbell Hall, Berkeley, CA, USA}

\author{Statia Luszcz-Cook}
\affiliation{University of Columbia, Astronomy Department, New York, USA}

\author{Katherine de Kleer}
\affiliation{California Institute of Technology, 1200 East California Boulevard, Pasadena, CA, USA}

\author{Arielle Moullet}
\affiliation{SOFIA/USRA, NASA Ames Building N232, Moffett Field, CA 94035, USA}

\author{Patricio M Rojo}
\affiliation{Universidad de Chile, Departamento de Astronomia, Casilla 36-D, Santiago, Chile}



\begin{abstract}

We present the first comprehensive study of NaCl and KCl gases in Io's atmosphere in order to investigate their characteristics, and to infer properties of Io's volcanoes and subsurface magma chambers. In this work, we compile all past spectral line observations of NaCl and KCl in Io’s atmosphere from the Atacama Large Millimeter/submillimeter Array (ALMA) and use atmospheric models to constrain the physical properties of the gases on several dates between 2012 and 2018. NaCl and KCl appear to be largely spatially confined and for observations with high spectral resolution, the temperatures are high ($\sim$500-1000 K), implying a volcanic origin. The ratio of NaCl:KCl was found to be $\sim$5-6 in June 2015 and $\sim$3.5-10 in June 2016, which is consistent with predictions based on observations of Io’s extended atmosphere, and less than half the Na/K ratio in chondrites. Assuming these gases are volcanic in origin, these ratios imply a magma temperature of $\sim$1300 K, such that the magma will preferentially outgas KCl over NaCl.

\end{abstract}



\section{Introduction} \label{sec:intro}

Jupiter’s moon Io, the innermost of the Galilean moons, is the most volcanically active body in our solar system due to intense tidal forces from its 4:2:1 orbital resonance with Europa and Ganymede. Strong internal heating from tidal forcing on Io results in extreme volcanic activity – there are over 400 known volcanoes on Io’s surface, approximately 150 of which had thermal activity detected during the Galileo mission \citep{lopes2004lava}. Io's atmosphere is primarily sulfur dioxide (SO\textsubscript{2}), thin and tenuous \citep{lellouch1990io}, and is replenished both by plume outgassing from its many active volcanoes, as well as sublimation of surface frost deposits \citep{lellouch2007io, tsang2016collapse, de2020alma}. As such, Io’s atmosphere is not uniform across its surface, but rather is highly spatially localized around active volcanoes and over surface frost deposits at latitudes $<|30-40|\degree$ (e.g., \citep{feaga2009io}; \citep{de20212020}).

Plume outgassing on Io is primarily SO\textsubscript{2}, with trace gases SO, NaCl, and KCl \citep{lellouch2007io, de20212020}. NaCl was first observed in Io’s atmosphere by \citeauthor{lellouch2003volcanically} (\citeyear{lellouch2003volcanically}), while KCl was first detected by \citeauthor{moullet2013exploring} (\citeyear{moullet2013exploring}). Long before NaCl was detected in Io’s atmosphere, its presence was predicted due to the strong detection of sodium in Io’s neutral clouds \citep{brown1974optical}, coupled to the ubiquity of alkali and chlorine outgassing in terrestrial silicate volcanism \citep{pennisi1988fractionation, symonds1992origin, symonds1994}. The origin of NaCl and KCl in Io’s atmosphere is likely volcanic outgassing, though sputtering of materials on Io’s surface may also be a source, which is the case for the atmospheres of the Moon and Mercury \citep{hunten1988mercury, killen1999surface}. Sputtering as a primary source of these materials, however, would be limited to areas where the atmospheric column density is $\le10\textsuperscript{16}$ cm\textsuperscript{-2}, such as at high latitudes or when Io is in Jupiter’s shadow and its SO\textsubscript{2} atmosphere has collapsed \citep{lanzerotti1982laboratory, de20212020}. In the case where these gases are indeed volcanic in origin, their ratio of abundances can be used to indirectly infer the temperature and composition of the magma chambers from which they originate \citep{fegley2000chemistry}, something which is at present poorly constrained.

No comprehensive studies correlating observations with models of NaCl and KCl in Io’s atmosphere have been carried out – photochemical models suggest that volcanically-outgassed NaCl and KCl would be removed from Io’s atmosphere in $\sim$3 hours \citep{moses2002alkali}, while observations of NaCl over a four-month period from 2016-2017 indicate that NaCl is remarkably stable over long time periods \citep{roth2020attempt}. In this work, we compile all past spectral line observations of NaCl and KCl in Io’s atmosphere from the Atacama Large Millimeter/submillimeter Array (ALMA) and use atmospheric models to constrain the physical properties of the gases (and Io’s putative subsurface magma chambers). Each observation was reduced and analyzed using identical spectral line modeling methods so that they may be assessed individually as well as jointly, in order to piece together an understanding of the nature of these gases in Io’s atmosphere. This paper has several main components: Section \ref{sec:obs} details the observations and data reduction, Section \ref{sec:model} describes the spectral line modeling used to analyze the data, Section \ref{sec:results} presents results from this analysis including the physical properties inferred from our models, Section \ref{sec:disc} discusses the common themes, interesting findings, anomalies, and remaining questions, and Section \ref{sec:conc} summarizes the main conclusions of the paper.

\begin{figure}[t!]
\plotone{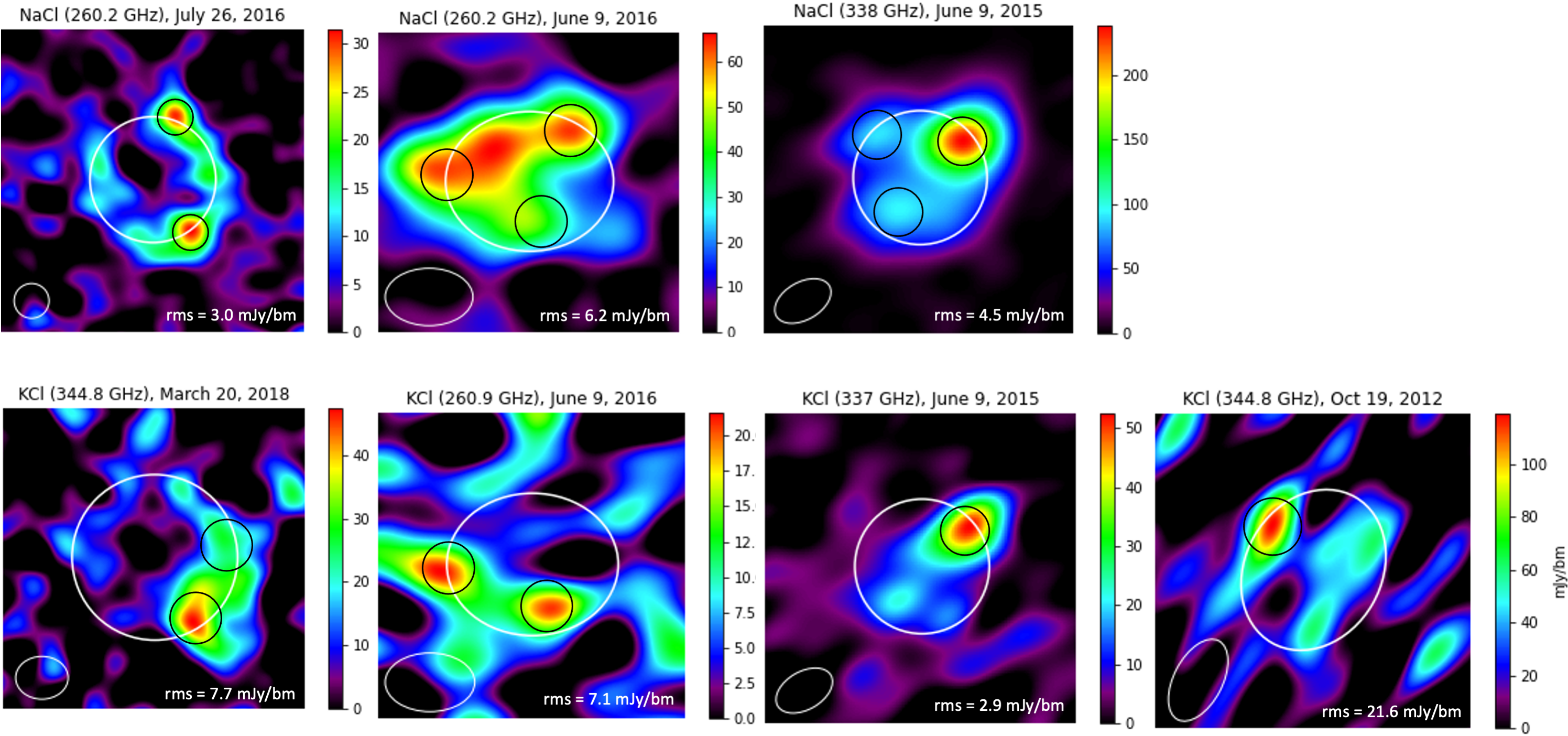}
\caption{All ALMA images used in this study, top row is NaCl lines, bottom row is KCl lines. Middle columns are dates where NaCl and KCl were observed concurrently. In these images (as well as all images in this paper), Io North is up, bottom left white circle is the HPBW of the synthesized beam (spatial resolution), center white circle is the outline of the disk of Io, taken from continuum maps of Io observed simultaneously, and black circles are the apertures analyzed in this study. Identification of each aperture (i.e. apertures 1, 2, etc.) is summarized in Figs. \ref{fig:appendix_nacl} and \ref{fig:appendix_kcl}). The rms error for each image, calculated from emission off Io's disk, is noted in the bottom right corner of each image} The radius of the apertures was determined by the smaller axis of the HPBW for each image (see Table \ref{tab:obs}). \label{fig:images}
\end{figure}

\begin{deluxetable*}{cccccc}
\tablenum{1}
\tablecaption{Spectral Line Species and Frequencies Observed\label{tab:lines}}
\tablewidth{0pt}
\tablehead{
\colhead{Species} & \colhead{Dates Observed} & \colhead{Frequency} & \colhead{Wavelength} &
\colhead{Line Strength} & \colhead{E (low)} \\
\nocolhead{Species} & \nocolhead{Species} & \colhead{(GHz)} & \colhead{(mm)} &
\colhead{(cm\textsuperscript{-1}/mol/cm\textsuperscript{2})} & \colhead{(cm\textsuperscript{-1})}
}
\startdata
KCl	& Oct 17 2012, Oct 19 2012, Mar 20 2018, Sept 2 2018, Sept 11 2018 & 344.820 & 0.869 & 2.235E-19 & 253.489 \\
NaCl & June 9 2015 & 338.021 & 0.887 & 2.747E-19 & 141.082 \\
KCl & June 9 2015 & 337.208 & 0.889 & 2.207E-19 & 242.241 \\
KCl & June 9 2016 & 260.916 & 1.149 & 1.647E-19 & 143.748 \\
NaCl & June 9 2016, July 26 2016 & 260.223 & 1.152 & 1.668E-19 & 82.510 \\
\enddata
\end{deluxetable*}

\section{Observations \& Data Reduction} \label{sec:obs}
\begin{wrapfigure}{r}{0.35\textwidth}
  \begin{center}
    \includegraphics[width=0.35\textwidth]{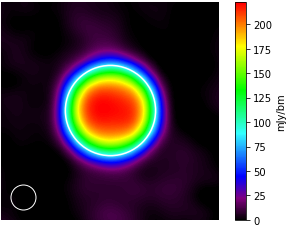}
  \end{center}
  \caption{Example continuum image from July 26, 2016, where the lower left white circle is the HPBW and the white circle in the center of the image is determined to be the edge of the disk of Io, which is then used to determine Io's size and location in the corresponding spectral line images.}
  \label{fig:continuum}
\end{wrapfigure}
Data were obtained on eight dates from 2012-2018 with the Atacama Large Millimeter/submillimeter Array (ALMA). All data are publicly available on the ALMA archive - in some cases, NaCl and/or KCl were the primary observational goals, and in some they were lines added to observations whose goals were primarily to observe other atmospheric gases (such as SO$_2$). Observations were made in Bands 6 and 7, which have central frequencies of 233 GHz (1.3 mm) and 343.5 GHz (0.87 mm) respectively. Several transitions of NaCl and KCl were observed, as detailed in Table \ref{tab:lines}. In two datasets, both NaCl and KCl were detected (June 9 2015, June 9 2016), in one both were targeted but only NaCl was detected (July 26 2016), and in two only a single line of KCl was targeted and detected (Oct 19 2012, March 20 2018), resulting in three total detections of NaCl and four detections of KCl. KCl was also targeted on October 17 2012, and September 2 and 11 2018, but not detected. SO$_2$ was observed on all dates where NaCl and KCl were observed. 

The spatial resolution varied greatly from observation to observation – in most cases, the half-power beam width (HPBW) is large and asymmetrical (see Figure \ref{fig:images}, bottom left corner in each image; Table \ref{tab:obs}). In the observation with the highest spatial resolution, $\sim$4 beams across the disk of Io were obtained. For each observation, all scans on Io have been combined to create a single spectral-line data cube per line per observation. In three datasets (July 26 2016, September 2 and 11 2018), Io went from eclipse into sunlight during the time of observation, and in one (March 20 2018) Io went from sunlight to eclipse – these scans have been combined in this analysis, since the eclipse had no apparent effect on the observed NaCl nor KCl emission when imaged separately. Several of the observations used in this study have been previously published, as shown in Table \ref{tab:obs}.
\begin{deluxetable*}{ccccccc}[b]
\tablenum{2}
\tablecaption{Observation Details\label{tab:obs}}
\tablewidth{0pt}
\tablehead{
\colhead{Date \& Time (UT)} & \colhead{Sub-Earth} & \colhead{Sub-Earth} & \colhead{Io diameter} & \colhead{HPBW} & \colhead{Resolution} & \colhead{ALMA Program,}\\
\colhead{(month/day/yr hr:m:s)} & \colhead{Longitude} & \colhead{Latitude} & \colhead{(arcsec)} & \colhead{(Bmax, Bmin, PA)} & \colhead{(kHz)} & \colhead{Publication}\\
\colhead{} & \colhead{(deg (W))} & \colhead{(deg (N))} & \colhead{} & \colhead{} & \colhead{} & \colhead{}\\
}
\startdata
Oct 17 2012 09:42:31-10:33:00 & 62.488 & 3.115 & 1.156 & 0.784”, 0.406”, -29.296$^{\circ}$ & 61.0 & 2011.0.00779.S \citep{moullet2015exploring} \\
Oct 19 2012 09:39:29-10:28:08 & 193.741 & 3.120 & 1.165 & 0.846”, 0.466”, -27.883$^{\circ}$ & 61.0 & 2011.0.00779.S \citep{moullet2015exploring} \\
June 9 2015 21:30:33-23:20:54 & 281.054 & -0.153 & 0.867 & 0.351”, 0.288”, -29.627$^{\circ}$ & 61.0 & 2012.1.00853.S \\
June 9 2016 01:14:54-02:17:07 & 24.587 & -1.550 & 0.931 & 0.585”, 0.358”, -59.54$^{\circ}$ & 245.6 & 2015.1.00995.S \\
July 26 2016 19:13:09-20:16:13 & 13.882 & -1.671 & 0.828 & 0.237”, 0.206”, -88.52$^{\circ}$ & 245.6 & 2015.1.00995.S \\
Mar 20 2018 10:54:43-11:13:53 & 343.264 & -3.405 & 1.058 & 0.328”, 0.284”, -83.470$^{\circ}$ & 144.0 & 2017.1.00670.S \citep{de2020alma}\\	
Sept 2 2018 21:46:21-22:40:08 & 19.472 & -2.957 & 0.885 & 0.30", 0.30", 0$^{\circ}$ & 144.0 & 2017.1.00670.S \citep{de2020alma}\\
Sept 11 2018 17:36:12-18:41:56 & 13.381 & -2.948 & 0.867 & 0.22", 0.22", 0$^{\circ}$ & 144.0 & 2017.1.00670.S \citep{de2020alma}\\
\enddata
\end{deluxetable*}

\begin{figure}[h]
\includegraphics[width=\textwidth]{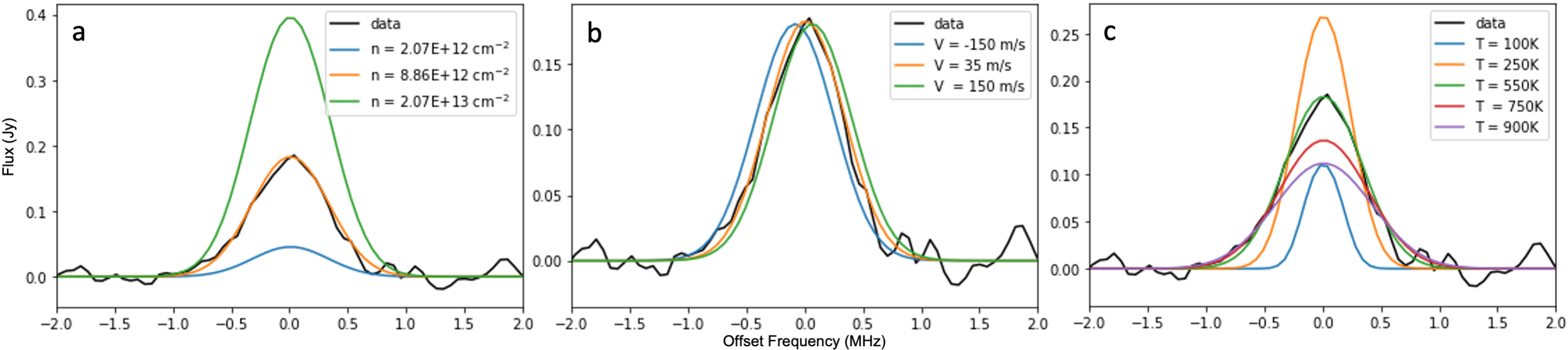}
\caption{Plots demonstrating the effect of varying column density (a), velocity (b), and temperature (c) on output spectral line profiles compared to observed line profile of NaCl aperture 1 on June 9 2015.  \label{fig:varying}}
\end{figure}

CASA version 5.4.0-68 was used for all steps of the data reduction and imaging. The raw data were calibrated and flagged via the ALMA pipeline, after which Io was split off into a separate calibrated measurement set (MS). An ephemeris file with Io’s position and velocity at a 1-minute time interval was attached to each MS to ensure Io’s tracking at high precision, which is necessary to distinguish gas velocity effects and get the right shape of the velocity profile in Io’s frame of motion. Io’s continuum was modeled as a limb-darkened disk – the limb-darkening coefficient and brightness temperature were adjusted to fit each individual observation, with limb-darkening generally being asymmetrical about the poles and equator (a value of q $\sim$ 0.3 North-South and q $\sim$ 0.1 East-West, where brightness falls off at higher emission angles ($\theta$) as a function of $\cos\textsuperscript{q}(\theta)$)). Peak brightness temperatures ranged from 80-100 K. This disk was used to self-calibrate the data, and to clean (i.e., deconvolve) the continuum map. Before imaging the line emission in each MS, the continuum emission was subtracted using CASA’s UVCONTSUB (see example continuum image Figure \ref{fig:continuum}), leaving only spectral line emission of each species (NaCl or KCl). Continuum images were used to determine the edge of the disk of Io for each spectral line image (see central white circles in Figure \ref{fig:images}). Line-averaged images of Io were created by averaging emission over 0.4 km/s centered at each spectral line’s rest frequency, except for June 9 2016 – these images are centered at -0.1 km/s from the line center for NaCl and -0.5 km/s from the line center for KCl, because these were the centers of the observed line profiles. The final line center maps are shown in Figure \ref{fig:images}. In order to model the spectral line profiles, image data cubes were created at a frequency resolution of 61 or 122 kHz (see Table \ref{tab:obs}), which allows one to construct a spectral line profile at any position on the disk. All images were created using uniform weighting to optimize for spatial resolution and cleaned using the Clark or Briggs algorithm in CASA’s TCLEAN routine. Line spectra were created from the data cubes by determining the flux density for each aperture (see apertures in Figure \ref{fig:images}) at each frequency - the line profile is constructed from the total flux density within each aperture at each frequency of the data cube. Examples of our spectral line profiles for NaCl aperture 1 from June 9 2015 are shown in Figure \ref{fig:varying}.

\section{Spectral Line Modeling} \label{sec:model}

Models of the spectral line data cubes were created using a radiative transfer code which integrates the equation of radiative transfer over a model of Io's atmosphere consisting of overlying plane-parallel layers. This code was originally developed to model CO lines in Neptune’s atmosphere \citep{luszcz2013constraining} and was subsequently adapted to Io’s atmosphere \citep{de2020alma} - the code used for this analysis is the same as what is used in \citet{de2020alma}. The models generated by this radiative transfer code assume that Io's atmospheric gases are in local thermodynamic equilibrium and hydrostatic equilibrium. Pixel size, beam size, spectral resolution, and frequency in each model were matched to the corresponding observations, while temperature, column density, and velocity of the atmospheric gases were varied, and the chi-square ($\chi\textsuperscript{2}$) goodness-of-fit was calculated to find which model spectral line output best fits the observed spectral line profile and infer reasonable constraints on the free parameters.
To demonstrate how the free parameters (temperature, column density, and velocity) affect the model spectral line output, in Figure \ref{fig:varying} we show the effect that each individual parameter has on the output line profile. When column abundance is increased and all other parameters are held constant, the peak of the line profile increases (see Figure \ref{fig:varying}a), due to an increased number of molecules in the system which emit photons at the central line frequency. Changing the gas velocity shifts the central frequency of the line profile (shifting to higher frequencies for negative velocity and lower frequencies for positive velocities, see Figure \ref{fig:varying}b). When the gas temperature is increased, the result is an increased line width due to Doppler broadening, which is the result of increased thermal motion, according to the following equation:

\begin{equation}
\triangle\nu = \frac{\nu\textsubscript{0}}{c}\sqrt{\frac{2kT}{m}}
\end{equation}
\begin{wrapfigure}{r}{0.5\textwidth}
  \begin{center}
    \includegraphics[width=0.55\textwidth]{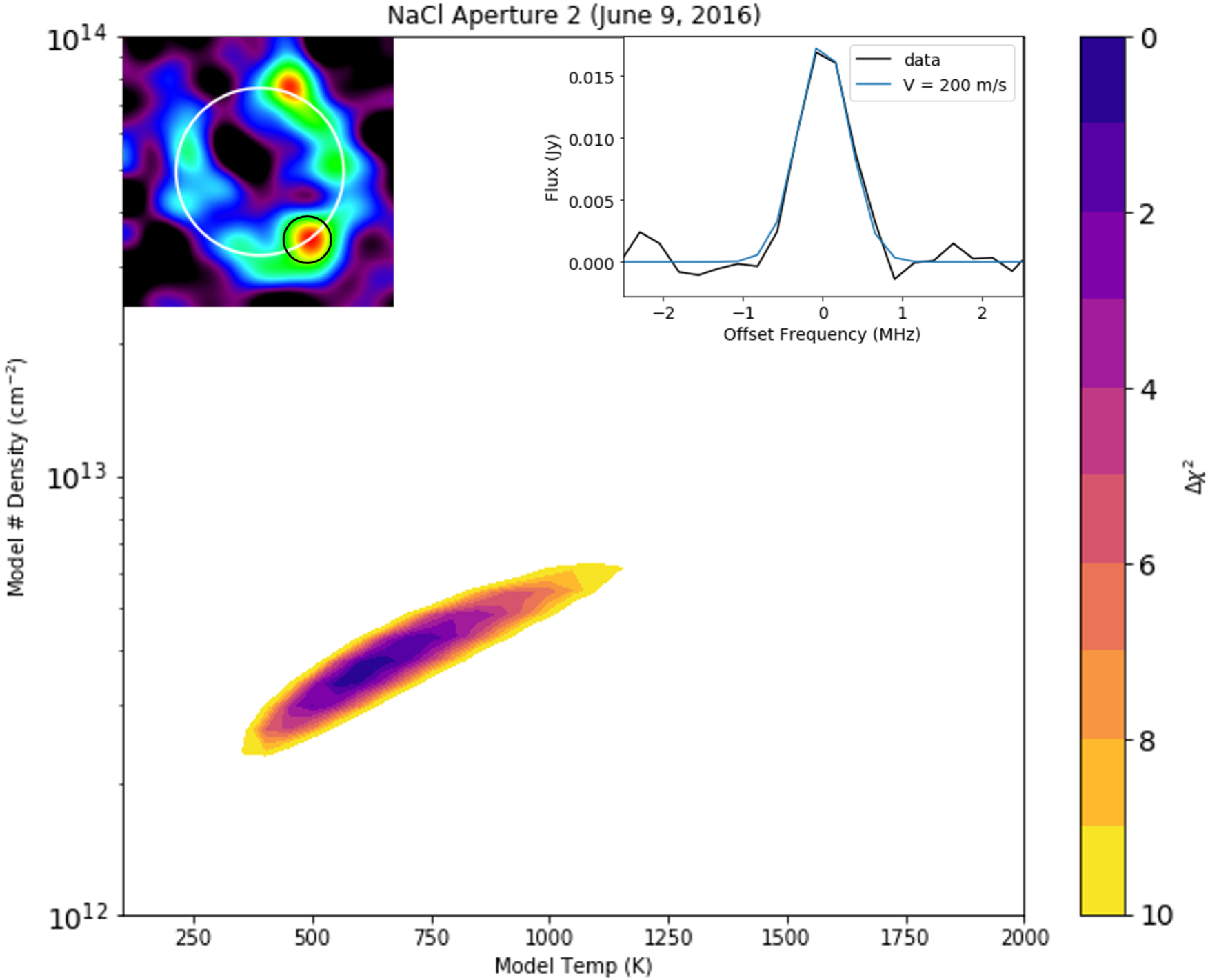}
  \end{center}
  \caption{Example plot of $\Delta\chi^2$ model fits to NaCl (260.2 GHz) observed on July 26, 2016 for temperatures ranging from 0 - 2000 K, column densities ranging from 10\textsuperscript{12}-10\textsuperscript{14} cm\textsuperscript{-2}, and a velocity of 120 m/s. Top left inset image indicates the aperture modeled, top right inset shows the line profile with the best fit line model. These plots were made for each aperture analyzed in this work, and are shown in the Appendix Figures \ref{fig:appendix_nacl} and \ref{fig:appendix_kcl}.}
  \label{fig:example}
\end{wrapfigure}
where $\triangle\nu$ is the width of the line profile at half-maximum, $\nu\textsubscript{0}$ is the central frequency of the line, $c$ is the speed of light, $k$ is Boltzmann’s constant, $T$ is the temperature of the gas, and $m$ is the mass of the molecule.

As such, temperature is the factor which determines the width of the line profile – increasing temperature leads to a larger velocity distribution of particles ($\triangle\nu \sim \sqrt{T}$), and therefore a broader line profile (see Figure \ref{fig:varying}c). Increasing temperature also affects the peak of the line profile since it increases the overall energy in the system. However, since higher temperatures lead to broader velocity distributions of particles, at some point increasing temperature will result in a decrease of the peak of the line profile because there are fewer molecules to emit energy at the line center since the molecules are moving around and emitting at nearby frequencies. As a result, when all other parameters are held constant, increasing temperature results in a consistently increasing line width, while the line peak will either increase or decrease depending on the number of emitting molecules in the system. While the effect of each individual parameter cannot be completely disentangled, their behaviors are unique enough to infer constraints on their values for each spectral line observation - density determines the height of the peak of the line profile, velocity shifts the line profile, and temperature broadens the line profile.

\begin{figure}[b]
\includegraphics[width=\textwidth]{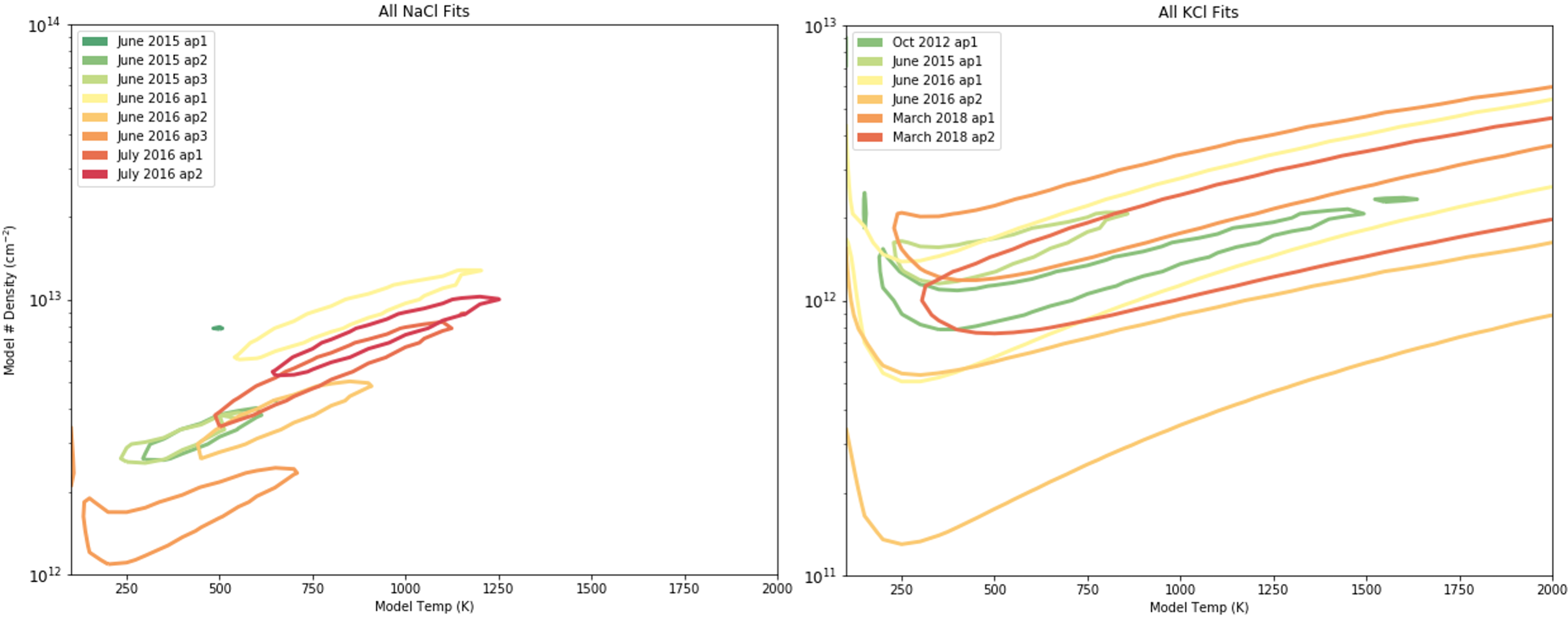}
\caption{Plots of $\Delta\chi^2=4.61$ contours for each aperture, NaCl apertures on the left and KCl apertures on the right. Detailed plots of each aperture are shown in Figure \ref{fig:appendix_nacl} and \ref{fig:appendix_kcl}. \label{fig:contours}}
\end{figure}

A grid of spectral line models was run for each observed line profile, where the temperature was varied from 100-2000 K in 50 K steps, and the column number density from 10\textsuperscript{12}-10\textsuperscript{14} cm\textsuperscript{-2} for NaCl and 10\textsuperscript{11}-10\textsuperscript{13} cm\textsuperscript{-2} for KCl in \begin{wraptable}{l}{0.4\textwidth}
\tablenum{3}
\caption{Velocities for Each Aperture}\label{tab:velocities}
\begin{tabular}{cc}\\\toprule  
Model Line & Velocity\\ \toprule
October 19, 2012, KCl, aperture 1 & -430 m/s \\
June 9, 2015, NaCl, aperture 1 & 35 m/s \\
June 9, 2015, KCl, aperture 1 & 25 m/s \\
June 9, 2015, NaCl, aperture 2 & -180 m/s \\
June 9, 2015, NaCl, aperture 3 & -250 m/s \\
June 9, 2016, NaCl, aperture 1 & -325 m/s \\
June 9, 2016, KCl, aperture 1 & 200 m/s \\
June 9, 2016, NaCl, aperture 2 & 120 m/s \\
June 9, 2016, KCl, aperture 2 & 150 m/s \\
June 9, 2016, NaCl, aperture 3 & 600 m/s \\
\end{tabular}
\end{wraptable} an equivalent number of logarithmically scaled steps. At each step through the model grid, a chi-square value was calculated to determine how well the output model line fits the observed spectral line profile:

\begin{equation}
\chi^2 = \sum_{i=1}^{N}\frac{(O_i - E_i)^2}{\sigma_i^2}
\end{equation}

where $N$ is the number of data points in the line profile, $O_i$ is the observed flux at each data point, $E_i$ is the model flux at each data point, and $\sigma$ is the root mean square (RMS) uncertainty.

After chi-square values were calculated for each point in the model grid, the minimum chi-square value was found and subtracted from each chi-square value in the grid:

\begin{equation}
\Delta\chi^2 = \chi^2 - \chi_{min}^2
\end{equation}

This created an output grid of $\Delta\chi^2$ values which essentially formed contours of constraints on temperature and density values for each observation (see example Figure \ref{fig:example}). This method can be used to determine the range of model parameter values which fit the data to a given confidence level (i.e., a value of $\Delta\chi^2 = 4.61$ gives a 90\% confidence level for a two parameter model fit \citep{avni1976energy, wall2012practical}).

\begin{wrapfigure}{r}{0.5\textwidth}
  \begin{center}
    \includegraphics[width=0.55\textwidth]{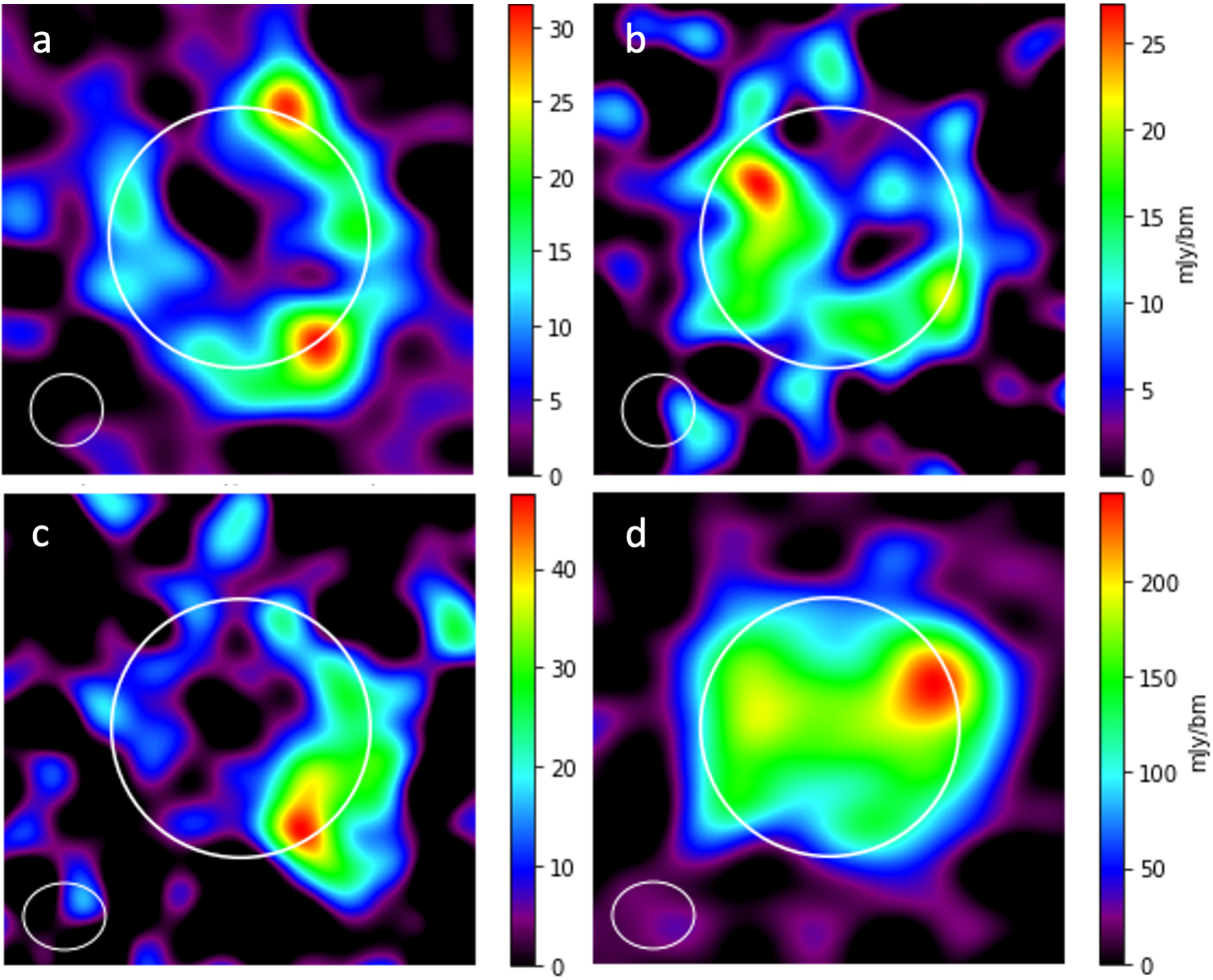}
  \end{center}
  \caption{a) NaCl emission (260.2 GHz) from July 26, 2016, b) SO\textsubscript{2} emission (262.2 GHz) in eclipse taken concurrently to a), c) KCl emission (344.8 GHz) from March 20, 2018, d) SO\textsubscript{2} emission (346.6 GHz) in eclipse taken concurrently to c). All maps were integrated over 0.4 km/s at the line center.}
  \label{fig:naclso2}
\end{wrapfigure}

\begin{figure}[h]
\includegraphics[width=\textwidth]{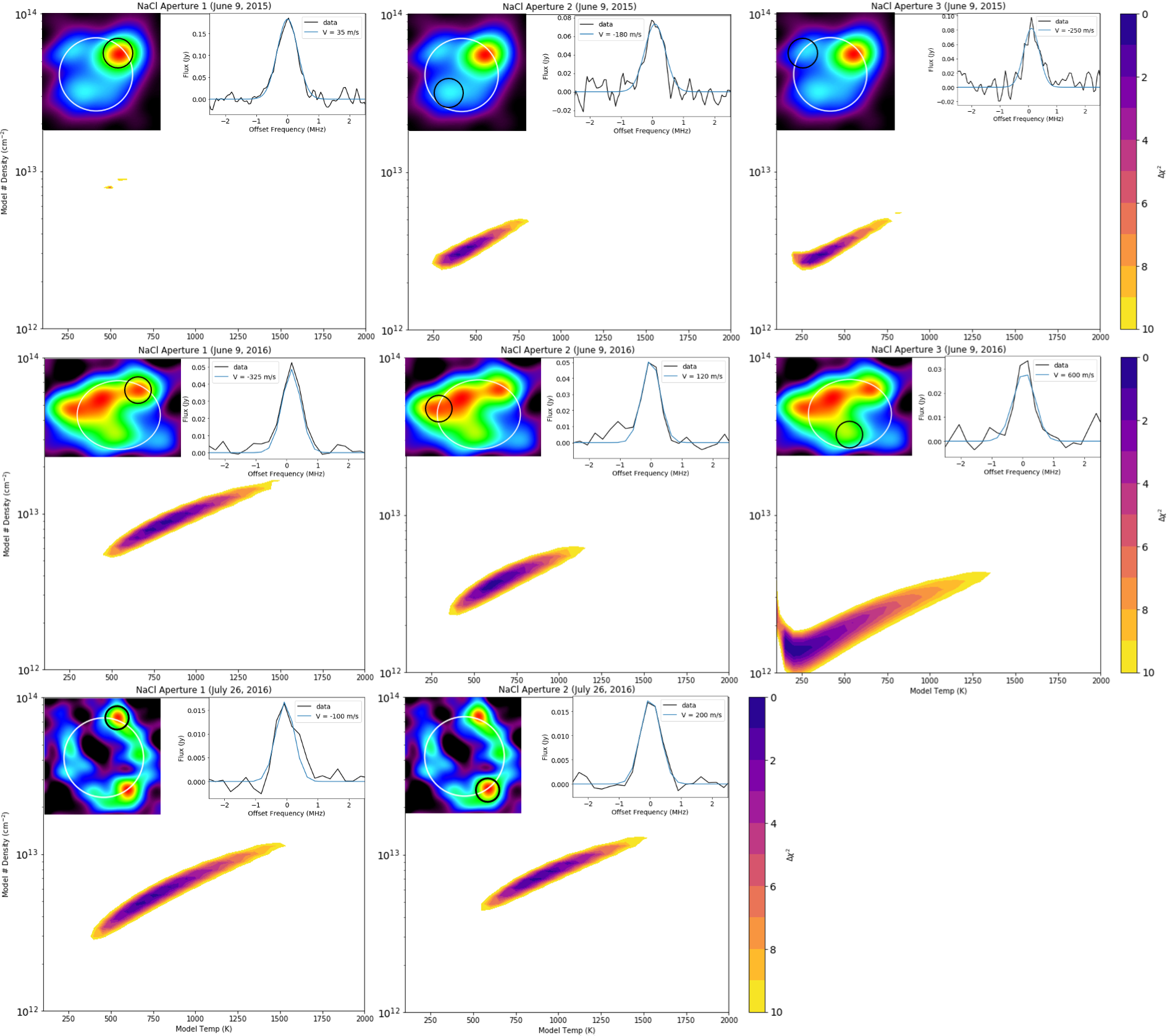}
\caption{Delta chi-square contour grids for all NaCl apertures analyzed in this study. Overlain on each plot is an image of the aperture and a smaller plot with the spectral line profile for the accompanying contour grid plot, along with the best fit output model line profile, labelled with the velocity at which the line profile is centered. \label{fig:appendix_nacl}}
\end{figure}

\begin{figure}[h]
\includegraphics[width=\textwidth]{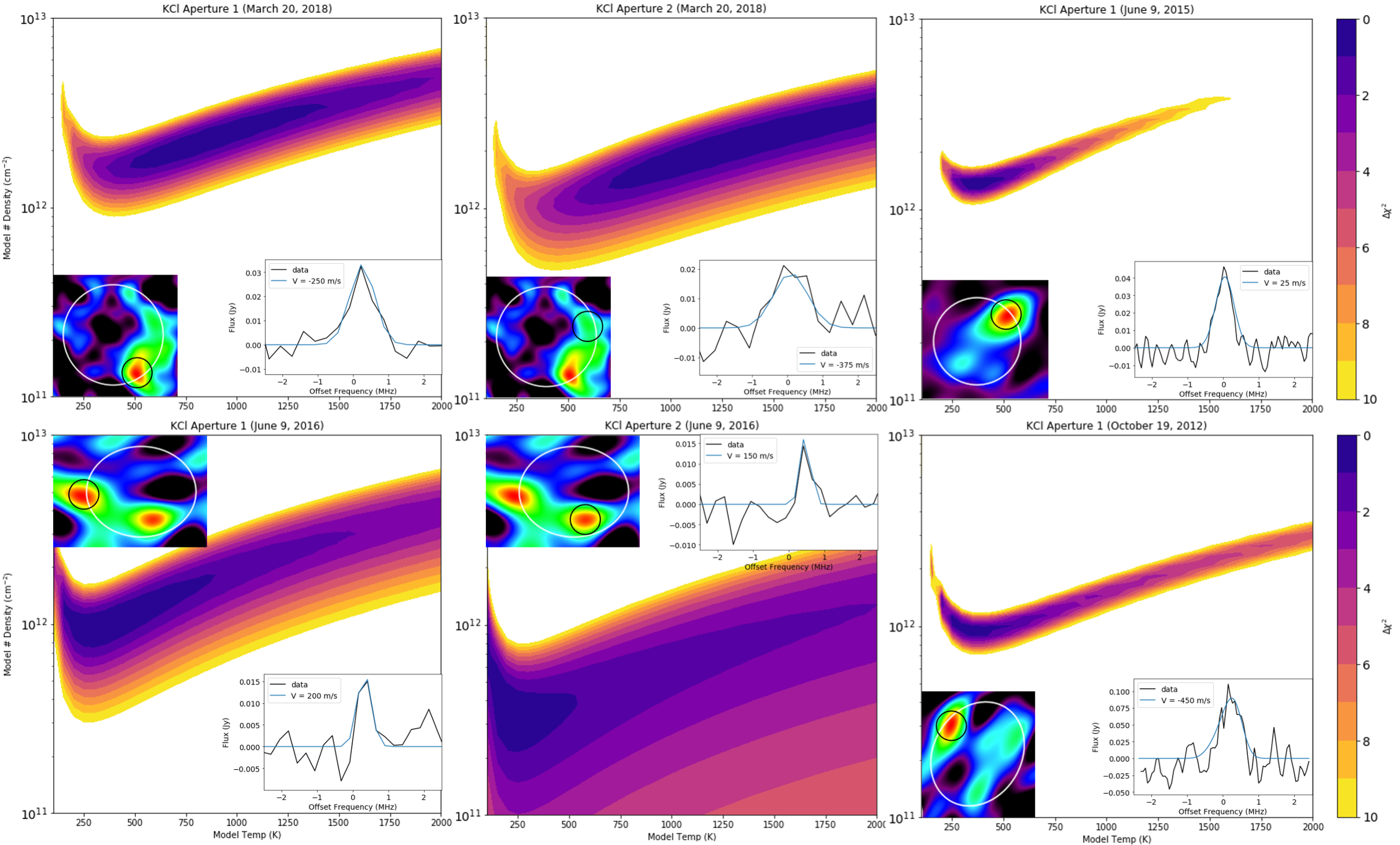}
\caption{Delta chi-square contour grids for all KCl apertures analyzed in this study. Overlain on each plot is an image of the aperture and a smaller plot with the spectral line profile for the accompanying contour grid plot, along with the best fit output model line profile, labelled with the velocity at which the line profile is centered. \label{fig:appendix_kcl}}
\end{figure}

Since the only effect of varying gas velocity in our model is shifting the line profile to higher or lower frequencies, this parameter was chosen before running the model grid by matching the peak of the model line profile to the peak of the observed line profile to ensure that the model line profile was centered on the observed line profile. These values are summarized in Table \ref{tab:velocities} for each aperture, which can be identified in detail in Figures \ref{fig:appendix_nacl} and \ref{fig:appendix_kcl}. Even large changes in velocity shift the line profile very minimally - see Figure \ref{fig:varying}b which shows a shift of +/-100 m/s on either side of the line profile. For most apertures, the spectral resolution is too low to allow for precise velocity measurements (see spectral resolutions in \ref{tab:obs}.

\section{Results} \label{sec:results}

\subsection{Constraints on NaCl and KCl Temperature and Column Density} \label{subsec:contours}
The primary finding of this project are constraints on model temperature and number density of NaCl and KCl gases for several ALMA spectral line observations throughout the years, which are shown in Figure \ref{fig:contours}. For all of these observations, NaCl and KCl were highly spatially localized around certain positions, so rather than model the physical parameters for these gases integrated over the disk, we instead chose smaller apertures at the emission maxima in each image (see Figure \ref{fig:images}). Aperture size for each observation was determined by the minor axis of the beam size in each image (see Bmin in Table \ref{tab:obs}).

The contours in Figure \ref{fig:contours} represent the boundary of model fits for which the delta chi-square value was less than or equal to 4.61 (90\% confidence). The size of the $\Delta\chi^2=4.61$ contour for each aperture also shows the relative quality of the data – data which had good spectral resolution and high signal to noise ratios resulted in smaller $\Delta\chi^2=4.61$ contours, meaning that a smaller range of values fit the line profile well. This is because observations with higher spectral resolution resulted in line profiles which had more data points to fill out the line profile in greater detail, while observations with lower spectral resolution resulted in line profiles whose shape was less well constrained. As such, plotting contours of all $\Delta\chi^2=4.61$ fits shows the results of all model fits while also representing the quality of the data used to determine the results.

\begin{wrapfigure}{l}{0.5\textwidth}
  \begin{center}
    \includegraphics[width=0.5\textwidth]{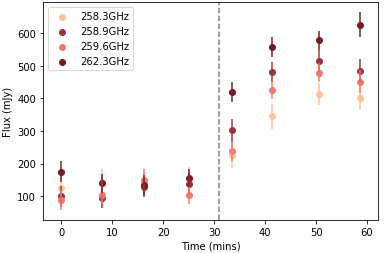}
  \end{center}
  \caption{Flux density vs. time for four different SO\textsubscript{2} transitions on July 26, 2016 when Io was observed moving from eclipse into sunlight. The flux densities have been determined from the center line emission maps of each $\sim$3-min scan (i.e., maps integrated over 0.4 km/s) by integrating the flux density on each map over the disk of Io. The gray dotted line represents the beginning of eclipse egress (when Io first moves into sunlight), after which SO\textsubscript{2} atmosphere reforms.}
  \label{fig:so2ecl}
\end{wrapfigure}

Model fits for column densities are reasonably well-constrained across all observations, likely because density is inferred through the line strength which is usually well-captured even in observations with low spatial or spectral resolution. However, model fits for temperature are generally poorly fit except for observations with high spectral resolution, since these observations have more data points which fill out the shape of the line profile more precisely. For the observations where temperature is well constrained, temperatures are relatively high $\sim$500-1000 K, suggesting a volcanic outgassing source.

For the dates where both NaCl and KCl were observed concurrently (June 2015 and June 2016), the ratio of NaCl:KCl gas abundance was calculated for a variety of model fits within the $\Delta\chi^2=4.61$ solution contour. This ratio has been determined for Io’s extended atmosphere (Na:K $\sim$ 7-13 \citep{brown2001potassium}) and for chondrites (Na:K $\sim$ 13-17, \citep{fegley2000chemistry}), which are assumed to be a proxy for Io’s interior ratio. However, this ratio had not been determined for Io’s atmosphere. We found that at all temperatures and densities within our $\Delta\chi^2=4.61$ contour fits, the NaCl:KCl ratio was $\sim$ 5-6 for the June 2015 data and $\sim$ 3.5-10 for the June 2016 data. This ratio was calculated using the possible range of NaCl and KCl column density values within the $\Delta\chi^2=4.61$ contour fits, assuming NaCl and KCl gases are the same temperature. These ratios are lower than what has been found for Io’s extended atmosphere and chondrites. Possible reasons for this difference are discussed in Section \ref{sec:disc}.

\subsection{NaCl/KCl and Sulfurous Compounds Are Not Colocated} \label{subsec:so2}

For each dataset in our analysis, at least one SO\textsubscript{2} transition was observed at the same time as NaCl/KCl, and two of the datasets involved the observation of SO\textsubscript{2} in eclipse (see Figure \ref{fig:naclso2}). Images of the SO\textsubscript{2} center line emission were created in the same way as our NaCl/KCl maps, i.e., the data are integrated over 0.4 km/s, centered at the center of the observed line profile. Image data cubes are based on the channel widths listed in Table \ref{tab:obs}. While in sunlight, SO\textsubscript{2} gas in Io’s atmosphere is sourced both from sublimation of SO\textsubscript{2} surface frost as well as volcanic outgassing \citep{spencer2005mid, lellouch2007io, lellouch2015detection}. However, once Io enters Jupiter’s shadow (a process which occurs once per Io day, i.e., roughly every 42 hours), its temperature drops below the vapor pressure point of SO\textsubscript{2}, resulting in a collapse of its SO\textsubscript{2} sublimation atmosphere \citep{saur2004relative, tsang2016collapse}. Io’s atmosphere reforms once it re-enters sunlight on a timescale of minutes \citep{de2020alma}. Figure \ref{fig:so2ecl} shows this process for SO\textsubscript{2} observations from July 26, 2016. As such, when SO\textsubscript{2} is observed while Io is in eclipse, one can assume that most observed emission is due to volcanic outgassing \citep{de2020alma}.

\begin{wrapfigure}{r}{0.5\textwidth}
  \begin{center}
    \includegraphics[width=0.55\textwidth]{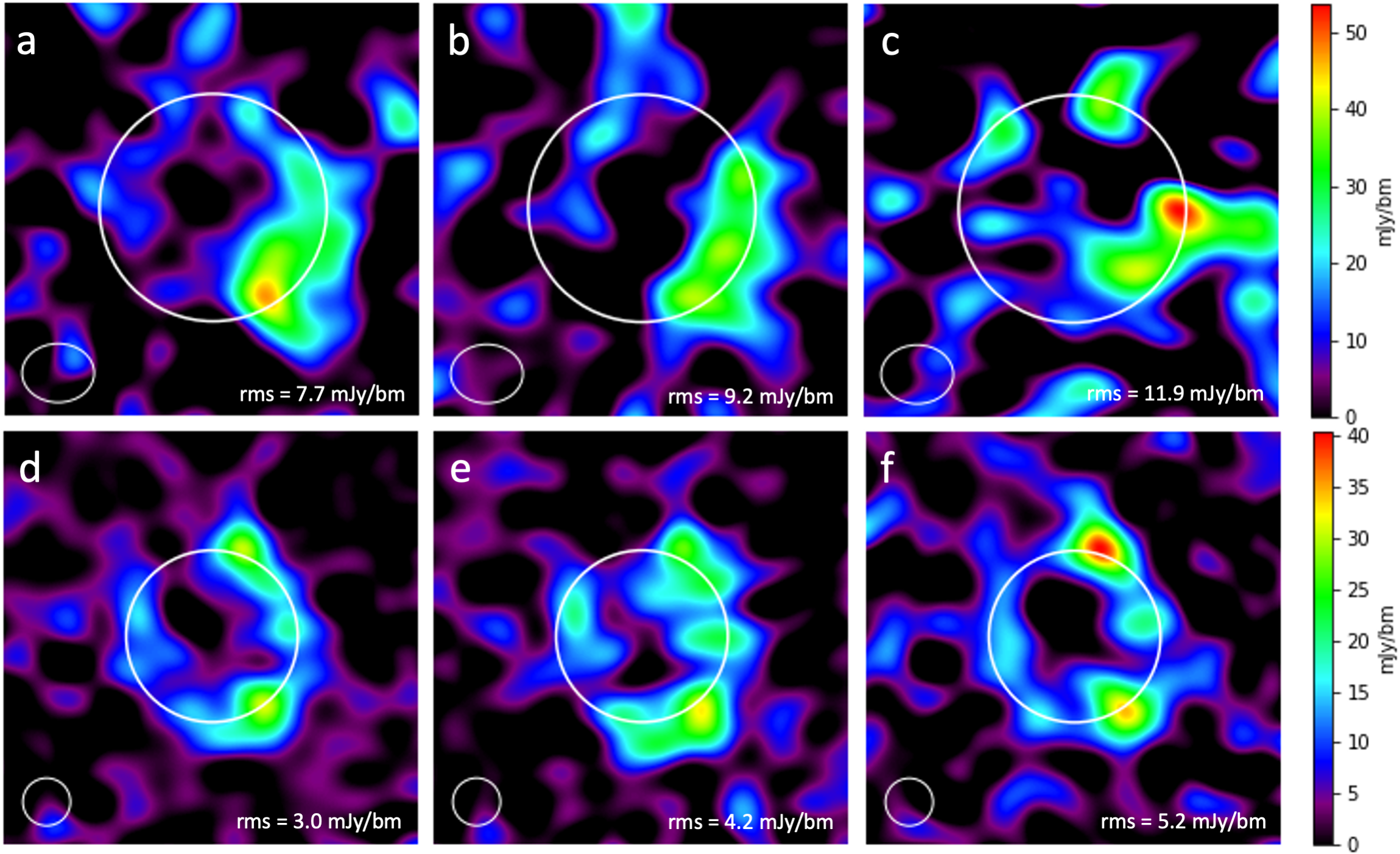}
  \end{center}
  \caption{Top row: KCl emission from March 20, 2018 imaged over all scans (a), and separated into sunlight (b) and eclipse (c) scans. Bottom row: NaCl emission from July 26, 2016 imaged over all scans (d), and separated into sunlight (e) and eclipse (f) scans. All images from the same dates are imaged on the same color scale for easy comparison, and rms error is noted in bottom corners of each image.}
  \label{fig:naclkcl_ecl}
\end{wrapfigure}

Figure \ref{fig:naclkcl_ecl} shows images from March 20, 2018 and July 26, 2016 when KCl and NaCl were imaged in eclipse and sunlight - it is also notable that NaCl and KCl showed no significant difference in emission between sunlight and eclipse, suggesting that Io's sulfurous atmospheric collapse and reformation has no effect on its alkali atmospheric gases.  While there are subtle differences between the sunlight and eclipse images, these variations are much lower than the rms error for each image. As such, when NaCl and KCl were imaged in sunlight and eclipse, these scans have been combined to boost the SNR. Assuming that NaCl and KCl are volcanic in origin, it is interesting to compare the locations of these gases with the location of concurrently observed volcanic SO\textsubscript{2} gas in the eclipse datasets. These two datasets (July 26, 2016 and March 20, 2018) observed just before eclipse egress and ingress, respectively, are almost at the same viewing geometry of Io (a difference of 5-10$^{\circ}$ in rotation).

\begin{figure}[b]
\includegraphics[width=\textwidth]{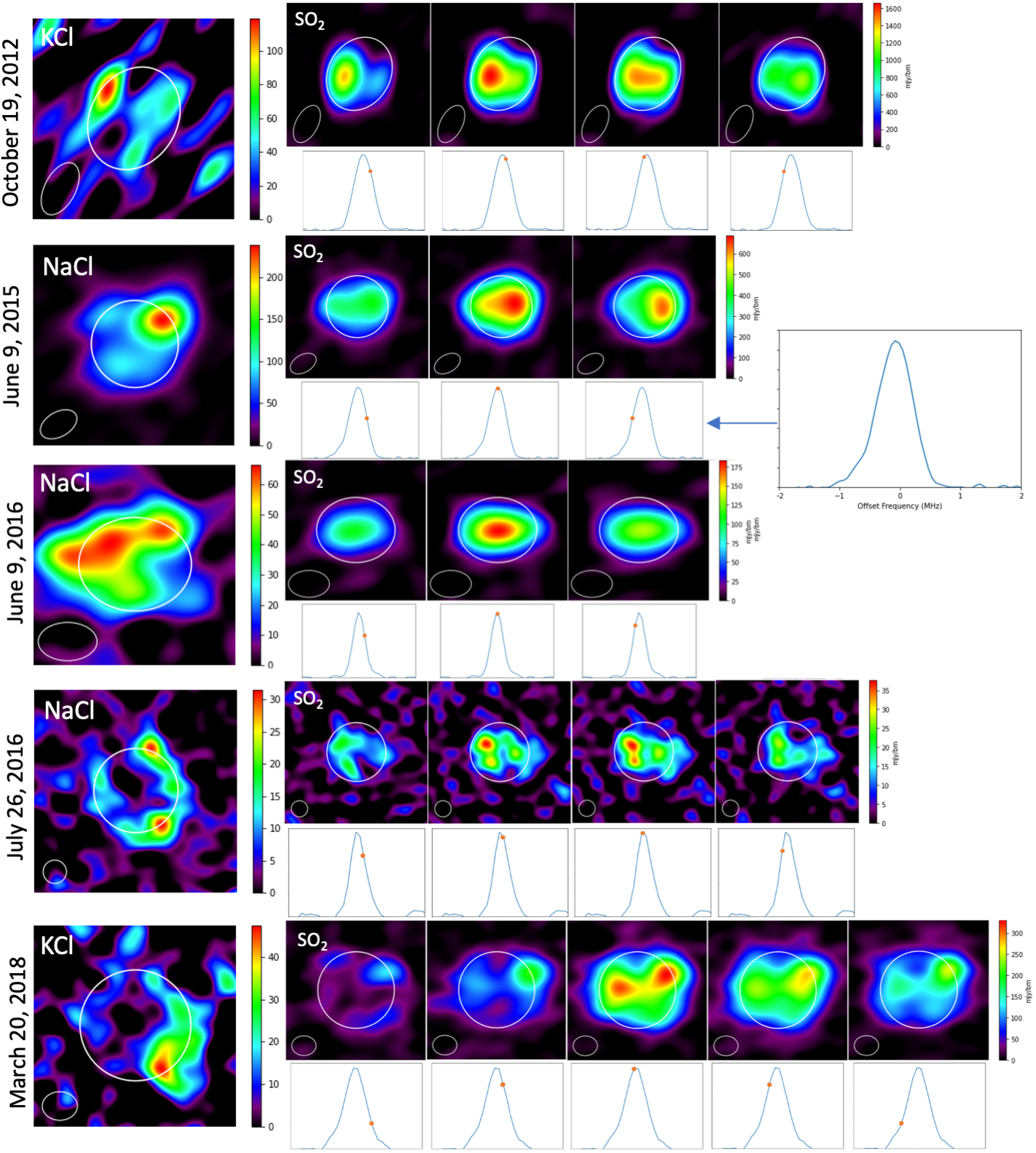}
\caption{Each row shows NaCl/KCl emission on far left for each date of observation in this study, followed by a series of images of SO\textsubscript{2} emission taken concurrently. For all images, all scans (sunlight and/or eclipse) are combined. SO\textsubscript{2} images from left to right show SO\textsubscript{2} emission across the emission line, where the leftmost image is emission on the blue-shifted side of the line wing, the center image is the emission at the center of the line, and the rightmost image is emission on the red-shifted side of the line wing. See \citet{de2020alma} for more detailed analysis. \label{fig:so2loc}}
\end{figure}

In the NaCl and KCl images for both dates, emission from these two trace species appear in similar areas – along the right (East) limb and at high eastern latitudes. Interestingly, it appears that the locations of peak volcanic SO\textsubscript{2} emission in these observations are not the same as the locations of peak NaCl and KCl emission observed concurrently. The SO\textsubscript{2} emission in eclipse also correlates to the SO\textsubscript{2} emission in the far line wings of the combined sunlight and eclipse images (see Figure \ref{fig:so2loc}) – line wing emission is due to SO\textsubscript{2} gas which is either at a high temperature or a high velocity, both suggesting volcanic emission \citep{de2020alma}. 

The presence of SO\textsubscript{2} gas at the same locations in eclipse as well as in the far line wings strongly suggests that this is volcanic emission, and in both cases does not spatially correlate with NaCl or KCl observed concurrently (see Figure \ref{fig:so2loc}), which we argue are also volcanically sourced. The lack of spatial correlation between these gases may suggest that the volcanic processes which produce SO\textsubscript{2} outgassing are not the same as those responsible for outgassing of NaCl and KCl, or that there may be some heterogeneity in the composition of Io’s subsurface magma chambers \citep{de2020alma}. Possible explanations are discussed in Section \ref{sec:disc}.

\subsection{Correlating Surface Features (Volcanoes) to NaCl/KCl Emission} \label{subsec:vol}

\begin{figure}[h]
\includegraphics[width=\textwidth]{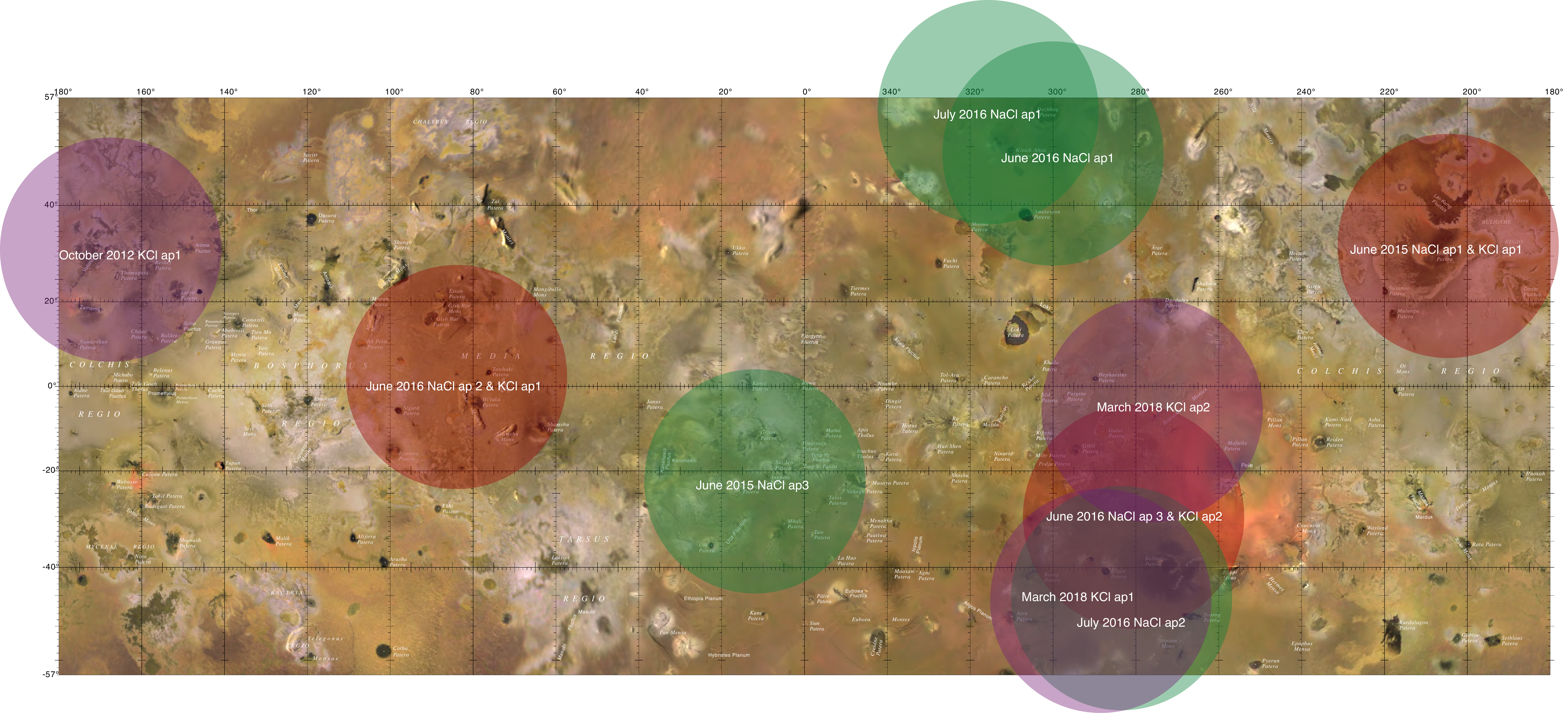}
\caption{Approximate locations of NaCl and KCl emission apertures from all observations. Purple circles are approximate KCl locations, green circles are approximate NaCl locations, and red circles are approximate locations where NaCl and KCl were observed concurrently. Base map from \citep{williams2011geologic}. \label{fig:iomap}}
\end{figure}

If the observed NaCl and KCl indeed result from volcanic outgassing, then it would only make sense that their locations would be correlated to known volcanoes on Io’s surface. Unfortunately, it is difficult to concretely identify the volcanic source for the observed emissions: 1) the spatial resolution of most of the observations is quite poor (at the highest spatial resolution, $\sim$4 beams across the disk of Io), and 2) Io has hundreds of volcanoes on its surface, many in close proximity. Emission locations should be taken with a grain of salt due to the large beam size in many of the observations, but it is useful to note areas where NaCl and KCl hotspot emission has been seen over several observations - a surface map indicating Io's volcanic features, overlaid with the KCl and NaCl detections described in this paper, is shown in Figure \ref{fig:iomap}.

\begin{figure}[h]
\includegraphics[width=\textwidth]{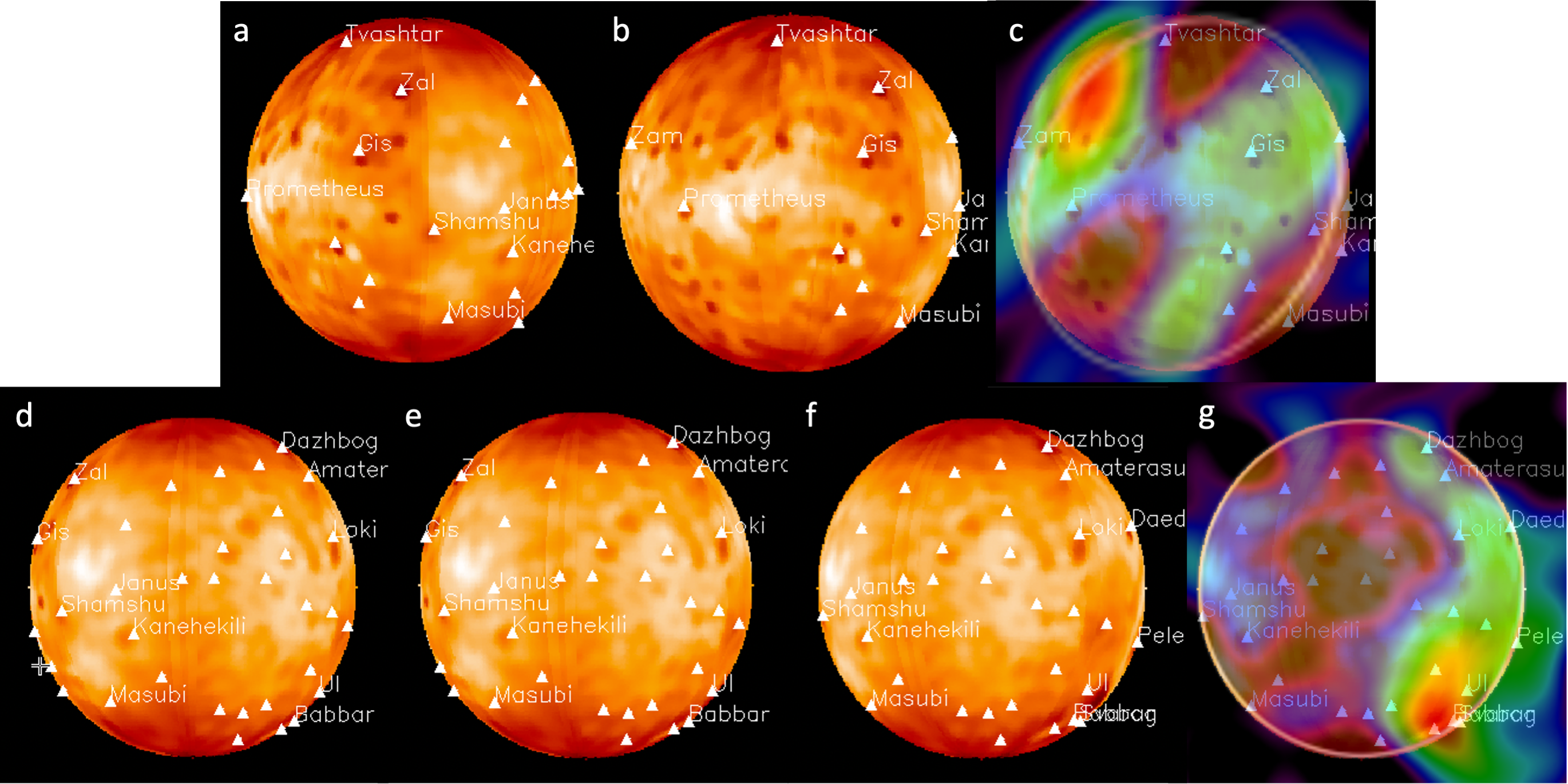}
\caption{Top row: a) View of Io on October 17, 2012, when observations taken of Io showed no KCl emission, b) View of Io on October 19, 2012 when KCl was detected, c) Same as (b) with observed KCl emission overlaid. Bottom row: Views of Io on d) September 2, 2018, and e) September 11, 2018, both of which showed no KCl emission, f) View of Io on March 20, 2018 when KCl was detected, g) Same as (f) with observed KCl emission overlaid. \label{fig:oct19}}
\end{figure}

It is also interesting to look at how non-detections are correlated to Io's orientation. Several dates analyzed in this work showed non-detections of KCl. The October 19 2012 KCl detection is an interesting case – for this dataset, Io was observed two nights prior (October 17 2012) and did not show KCl emission, despite having the same observational set-up and very similar views of Io's surface and similar levels of noise (see Figure \ref{fig:oct19}). Comparing the KCl emission maps to the map of volcanoes on Io for the times of observation show that from the first observation to the second, the primary change is that volcano Zamama had rotated into view. Zamama's location is near that where we see the observed KCl emission on October 19 (Figure \ref{fig:oct19}). While the spatial correlation of emission to Zamama may be spurious due to the spatial resolution and SNR of the images, the fact that KCl is not seen on October 17 but clearly seen on October 19 after Zamama has rotated into view is striking.

Zamama is a well-studied volcanic center on Io. Analysis of SSI and NIMS data from Galileo indicated silicate magma volcanism at a temperature of $\ge$1100 K \citep{davies1997temperature}. Surface flows observed on Zamama indicate that it is a Promethean-style volcano, which are characterized by long-term lava flows which are likely fed from shallow magma reservoirs \citep{davies2011variability}. SSI images revealed intermittent plume activity at Zamama \citep{geissler2008galileo}. 

No KCl emission was seen on September 2 and 11 2018, even though the emission frequencies were covered by the observations. A slightly different view of Io's surface was observed on March 20 2018, and KCl emission was seen on the eastern limb of Io over an area above Ulgen and Babbar Patera, which were barely on the edge of the limb in the September observations. This volcanic region was first identified in Voyager data \citep{pearl1982hot, mcewen1985volcanic}, whose initial analysis combined the two volcanoes into one hotspot due to their close proximity. Observations on March 8 2015 from the Large Binocular Telescope observed volcanic activity in this region \citep{de2021resolving}. 

NaCl and KCl were both observed on June 9 2015, and both emission hotspots are very strong and highly co-located with each other. The temperature range for NaCl was well-constrained to relatively high temperatures ($\sim$450-700 K), suggesting that these gases are volcanic in origin. Their emission appears to be directly located over Isum, though it should be noted that Zamama is also nearby. Isum has no known plume, and SO\textsubscript{2} was not observed over Isum at the time of observation \citep{de20212020}.

For the eclipse observations in Figure \ref{fig:naclso2}, the same side of Io is seen in both sets of images. The emission at high latitudes correlates to the area around Dazhbog Patera, though Amaterasu, Fuchi, Manua, and Kimich Ahau are all active volcanoes nearby. Plumes have been observed at Dazhbog Patera \citep{mcewen2002volcanic}, and recent lava flows characterize the geological makeup of the surface in the area \citep{williams2011geologic}. Along the eastern limb, emission is seen in both sets of images over the area around Babbar Patera and Ulgen Patera.

\section{Discussion} \label{sec:disc}

Several interesting questions emerge from our findings. The first is the apparent lack of correlation between volcanic SO\textsubscript{2} gas and NaCl/KCl emission. If both of these gases are from volcanic sources, then why do some volcanoes appear to outgas sulfurous compounds while some outgas alkali compounds? The first possible explanation is that for some volcanoes, SO\textsubscript{2} gas emission is the result of hot lavas vaporizing SO\textsubscript{2} surface frost deposits, rather than outgassing directly from the volcanic vent. In this scenario, SO\textsubscript{2} emission in eclipse would be constrained to the locations of SO\textsubscript{2} surface frost deposits, which are primarily found in Io’s equatorial band. This is consistent with our SO\textsubscript{2} eclipse and linewing observations (see Figures \ref{fig:so2ecl} and \ref{fig:so2loc}), which primarily show SO\textsubscript{2} emission at mid-latitudes.

Another possible explanation is that NaCl and KCl emission is confined to high temperature volcanoes, while SO\textsubscript{2} emission arises from magma chambers whose temperatures do not reach the boiling point of NaCl and KCl (1373 K for NaCl and 1173 K for KCl at Io surface pressures \citep{fegley2000chemistry}). If this were the case, SO\textsubscript{2} should be observed at all locations where NaCl and KCl are observed, while NaCl and KCl would be limited to only high temperature eruptions. Interestingly, NaCl and KCl are often observed at high latitudes, indicative of being sourced by deep mantle heating \citep{segatz1988tidal, de2016spatial, de20212020}. Perhaps at these latitudes, the temperature of the atmosphere is cold enough that SO\textsubscript{2} condenses out very quickly and is not detectable in our observations, although SO\textsubscript{2} emissions have been detected over energetic high latitude eruptions (e.g., Tvashtar, \citep{jessup2012characterizing}). It seems that there is not a clear explanation that satisfies all observations and open questions - observations with higher spatial resolution would help to clarify this conundrum.

Assuming the observed NaCl and KCl are indeed products of volcanic outgassing, then constraints on their abundance can be used to infer the temperature and composition of the magma chambers themselves \citep{fegley2000chemistry}. This relationship between observed atmospheric NaCl:KCl ratio and magma chamber conditions presents an exciting opportunity to constrain physical properties of Io’s magma chambers indirectly through observing plume outgassing emission. Magma thermochemical equilibrium calculations by \cite{fegley2000chemistry} suggest that the ratios found in our study correlate to a magma temperature around ~1300 K. 

Additionally, it is interesting to note that Io’s atmospheric NaCl:KCl ratio (NaCl:KCl $\sim$ 5-6, 3.5-10) is marginally lower than the ratio found for its extended atmosphere (Na:K $\sim$ 7-13, \citep{brown2001potassium}) and significantly lower than the ratio found in chondrites (Na:K $\sim$ 13-17, \citep{fegley2000chemistry}). The explanation for the difference between the ratio derived in this work and that of Io’s extended atmosphere can be explained by the analysis done by \citet{brown2001potassium}, who notes that his ratio is likely an upper limit on the Na:K ratio for Io’s surface and near-surface atmosphere, due to the difference in photoionization times for sodium and potassium (400 and 270 hours, respectively \citep{smyth1995theoretical}). Additionally, \citet{brown2001potassium} detected a high-speed jet (which they presumed to consist of material more recently ejected from Io’s surface) with a ratio $\sim$ 5-7, which is comparable to our ratios.

As for the difference in our ratio compared to the chondritic ratio, it is possible that this is the result of KCl’s lower condensation temperature (1173 K compared to 1373 for NaCl \citep{fegley2000chemistry}), which would result in KCl being preferentially outgassed in magmas whose temperatures are between these numbers, lowering the NaCl:KCl ratio in Io’s atmosphere as compared to the ratio in the magma \citep{de20212020}.

\section{Conclusions} \label{sec:conc}
We have presented the first comprehensive study of NaCl and KCl gases in Io's atmosphere. The primary findings of this paper include:
\begin{itemize}
  \item Temperature and column abundance constraints on NaCl and KCl spectral line emission from several dates over the period of 2012-2018
  \item NaCl and KCl were highly spatially confined and temperatures (where well-constrained) were high ($\sim$500-1000 K), suggesting that these compounds are volcanic in origin
  \item NaCl:KCl ratios of $\sim$5-6 in June 2015 and $\sim$3.5-10 in June 2016, which is consistent with (or perhaps marginally lower than) extended atmosphere calculations from \citet{brown2001potassium} and a factor of two lower than the chondritic Na:K ratio, implying a magma temperature of $\sim$1300 K
  \item Volcanic SO$_2$ emission and volcanic NaCl/KCl emission are not co-located, which could be explained by heterogeneity in Io's subsurface magma chambers or a difference in processes which preferentially outgas SO$_2$ vs. NaCl/KCl
  \item Several volcanoes (including Zamama and Isum Patera) were identified as possible sources of NaCl/KCl plumes
\end{itemize}

\subsection{Future Work} \label{subsec:future}
While the correlations between NaCl and KCl to known volcanoes on Io is intriguing, observations with higher spatial resolution are needed to correlate the emissions with a higher accuracy to a location of known volcanoes. Observations of these compounds at known active volcanoes could take place on a target-of-opportunity basis. Additionally, an atmospheric model including volcanic plumes in a sublimation atmosphere (such as those in \citep{zhang2003simulation}, \citeyear{zhang2004numerical}, \citep{mcdoniel2015three}) would be a great improvement over the modeling done for this paper, as the model used for this work assumes LTE and hydrostatic equilibrium, which is not ideal for active plume outgassing at extremely high temperatures.

\subsection{Acknowledgements} \label{subsec:acknow}

This paper makes use of the following ALMA data: ADS/JAO.ALMA\#2011.0.00779.S, ADS/JAO.ALMA\#2012.1.00853.S, ADS/JAO.ALMA\#2015.1.00995.S, and ADS/JAO.ALMA\#2017.1.00670.S. ALMA is a partnership of ESO (representing its member states), NSF (USA), and NINS (Japan), together with NRC (Canada), MOST and ASIAA (Taiwan), and KASI (Republic of Korea), in cooperation with the Republic of Chile. The Joint ALMA Observatory is operated by ESO, AUI/NRAO, and NAOJ. The National Radio Astronomy Observatory is a facility of the National Science Foundation operated under cooperative agreement by Associated Universities, Inc. All data can be downloaded from the ALMA Archive. This research was supported by the National Science Foundation.
Facility: ALMA.

\bibliography{sample631}{}
\bibliographystyle{aasjournal}

\subsection{Appendix}

\end{document}